\documentclass[journal=jacsat,manuscript=article]{achemso}

\usepackage{xcolor}

\author{Lily Yang}
\email{lilyy@kth.se}
\author{Stephan Steinhauer}
\affiliation{Department of Applied Physics, KTH Royal Institute of Technology, \\Albanova University Centre, SE-106 91 Stockholm, Sweden}
\author{Elia Strambini}
\affiliation{NEST, Istituto Nanoscienze-CNR and Scuola Normale Superiore, \\Piazza S. Silvestro 12, Pisa I-56127, Italy}
\author{Thomas Lettner}
\affiliation{Department of Applied Physics, KTH Royal Institute of Technology, \\Albanova University Centre, SE-106 91 Stockholm, Sweden}
\author{\\Lucas Schweickert}
\affiliation{Department of Applied Physics, KTH Royal Institute of Technology, \\Albanova University Centre, SE-106 91 Stockholm, Sweden}
\author{Marijn A.M. Versteegh}
\affiliation{Department of Applied Physics, KTH Royal Institute of Technology, \\Albanova University Centre, SE-106 91 Stockholm, Sweden}
\author{Francesco Giazotto}
\affiliation{NEST, Istituto Nanoscienze-CNR and Scuola Normale Superiore, \\Piazza S. Silvestro 12, Pisa I-56127, Italy}
\author{\\Valentina Zannier}
\affiliation{NEST, Istituto Nanoscienze-CNR and Scuola Normale Superiore, \\Piazza S. Silvestro 12, Pisa I-56127, Italy}
\author{Lucia Sorba}
\affiliation{NEST, Istituto Nanoscienze-CNR and Scuola Normale Superiore, \\Piazza S. Silvestro 12, Pisa I-56127, Italy}
\author{Dmitry Solenov}
\affiliation{Department of Physics, Saint Louis University, St. Louis, MO 63103, USA}

\title{Proximitized Josephson junctions in highly-doped \\InAs nanowires robust to optical illumination}


\begin{document}

\begin{abstract}
We have studied the effects of optical-frequency light on proximitized InAs/Al Josephson junctions based on highly \textit{n}-doped InAs nanowires at varying incident photon flux and at three different photon wavelengths.  The experimentally obtained IV curves were modeled using a shunted junction model which takes scattering at the contact interfaces into account. The Josephson junctions were found to be surprisingly robust, interacting with the incident radiation only through heating, whereas above the critical current our devices showed non-thermal effects resulting from photon exposure. Our work provides important guidelines for the co-integration of Josephson junctions alongside quantum photonic circuits and lays the foundation for future work on nanowire-based hybrid photon detectors.
\end{abstract}

Semiconductor-superconductor hybrid devices have attracted increasing attention in state-of-the-art quantum information processing. The interactions of these devices with electromagnetic radiation in the optical domain opens up exciting opportunities for both emerging technologies and fundamental science. Several optoelectronic device architectures have been proposed including entangled-photon pair sources\cite{Recher2010, Hayat2014}, Josephson lasers\cite{Godschalk2011} and photonic Bell-state analyzers\cite{Hayat2017}. Experimentally, enhanced photon generation in light-emitting diodes based on conventional epitaxially-grown semiconductor \textit{p-n} junctions contacted by superconducting leads have been demonstrated\cite{Sasakura2011}. Semiconducting nanowires constitute an important building block for hybrid devices relying on superconducting electrodes and the proximity effect, which was employed for the demonstration of gate-tunable Josephson junctions\cite{Doh2005}, quantum electron pumps\cite{Giazotto2011} and microwave quantum circuits\cite{deLange2015}. In particular, hybrid nanowire devices are extensively studied due to the promising prospects for topological quantum computing mediated by Majorana zero modes\cite{Lutchyn2018}. Proximitized nanowire junctions have been demonstrated in a variety of material systems, such as InAs\cite{Deng2016} or InSb \cite{Zhang2018} nanowires covered with epitaxial Al, InAs nanowires with Pb\cite{Paajaste2015} and Nb\cite{Guenel2012} contacts, InSb nanowires contacted by NbTiN leads\cite{Zhang2017}, InN nanowire-Nb junctions\cite{Frielinghaus2010}, PbS nanowires with PbIn electrodes \cite{Kim2017} and CdTe-HgTe core-shell nanowires in combination with Al contacts\cite{Hajer2019}. Surprisingly, the fundamental transport characteristics of proximitized nanowire junctions interacting with photons at optical wavelengths have remained unexplored despite the significant technological importance of the related phenomena, for instance in photon-qubit interfaces. The realization of large-scale quantum networks requires the combination of quantum hardware nodes and photonic platforms compatible with fiber-based telecommunication \cite{Wehner2018}, necessitating coherent interfaces between photons and qubits similar to those proposed for superconducting devices \cite{Andrews2014}, trapped ions\cite{Mehta2016} and solid-state spins\cite{Gao2015,Awschalom2018}. 

Optical wavelength photons can have a significan effect on the superconductivity in hybrid superconductor-semiconductor Josephson junctions.  Early experiments on light-sensitive semiconductor-superconductor junctions showed that CdS thin films between Pb or Sn electrodes could be switched to a Josephson state related to a persistent conductivity enhancement\cite{Giaever1968}. Furthermore, the interface barrier\cite{Akazaki2009} and the critical current\cite{Schaepers1999} of superconducting junctions on two-dimensional electron gases could be adjusted by light exposure. More recently, non-equilibrium effects of photoexcited carriers in Graphene-based Josephson junctions have been reported\cite{Tsumura2016}.  In particular, the electrical transport in low-bandgap semiconductors such as InAs have been shown to highly responsive to light\cite{Fang2016}, bringing to question whether hybrid superconducting devices based on such semiconductors can be operated in close proximity to photonic elements.

In this work, we present a comprehensive study on the electrical transport properties of highly \textit{n}-doped InAs nanowires proximitized by superconducting Al electrodes during exposure to light inside a dilution refrigerator. The nanowire Josephson junctions were exposed to laser illumination in the visible and infrared range, in particular at the three wavelengths 532\,nm, 790\,nm (around Rb transitions relevant for atomic quantum memories) and 1550\,nm (C-band telecommunication window). Experiments were performed at increasing incident photon flux and the obtained results were modeled using a shunted junction model that accounts for scattering at the semiconductor-superconductor interfaces. Using an independently measured temperature-dependent dataset, the IV characteristics were fitted to distinguish between thermal and non-thermal effects. Our results demonstrate the Josephson junctions' robustness to optical photon exposure, which has important implications for the implementation and operation of hybrid nanowire devices in integrated quantum photonic circuits.

\begin{figure}[t]
  \includegraphics[width=1\textwidth]{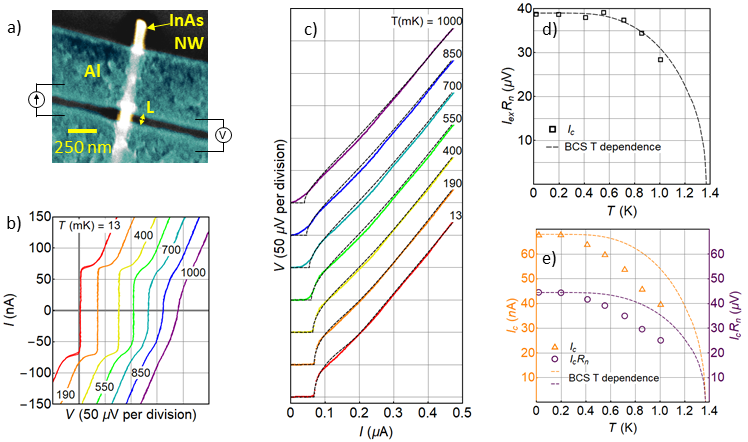}
  \caption{(a) Scanning electron microscopy image of n-doped InAs nanowire contacted by two Al leads separated by a gap $L$ of approximately 100\,nm. The four-point measurement configuration is schematically indicated. (b) IV characteristics obtained under DC current biasing at different cryostat temperatures (the sample was not exposed to photons during the measurements). (c) Experimental data fitted to a shunted junction model (black lines). (d) The product of excess current times normal-state resistance $I_{ex}R_n$ extracted from the model, showing good agreement with a BCS-type temperature dependence. (e) The critical current $I_c$ and the product of critical current times normal-state resistance $I_cR_n$ extracted from the model deviate from BCS-like behaviour, which indicates scattering in the junction.}
  \label{fgr1}
\end{figure}

Se-doped InAs nanowires were grown by Au-assisted chemical beam epitaxy. \cite{Gomes2015} The n-type nanowires  have an average diameter of 80 nm, length of 2.6 micrometer, and an electron concentration of about $1\times10^{18}\,cm^{-3}$. The nanowire-based devices were fabricated on SiO$_2$/Si substrates by electron beam lithography followed by magnetron sputtering of Al and a standard lift-off process. In situ etching with Ar ions was performed to remove the native oxide layer on the InAs nanowire surfaces. Details related to the fabrication process can be found in the Supporting Information. The InAs nanowires were contacted by two Al leads separated by a gap $L$ of approximately 100\,nm (Fig.\,\ref{fgr1}a), which forms the Josephson junction. The leads were used for both current biasing and voltage measurements in a four-point configuration. The transport characteristics were first measured as a function of temperature without photon exposure by heating the sample stage in the dilution refrigerator. The experimentally obtained IV curves (Fig.\,\ref{fgr1}b) show a gradual suppression of proximity-induced superconductivity in the InAs nanowire with increasing temperature. The slightly rounded transitions between the superconducting and the normal state without hysteresis is consistent with previous studies of InAs nanowire Josephson junctions with medium contact transparency\cite{Gharavi2017}; a more detailed discussion on the junction transparency will be presented below.

The IV characteristics of the InAs nanowire Josephson junctions can be described using a shunted junction model. A normal (non-coherent) current $I_n$, propagating via electronic excitations in the conductive nanowire, is added to a coherent Josephson supercurrent tunneling through the nanowire junction:\cite{AslamazovLarkinOvchinnikov}
\begin{eqnarray}\label{eq:I}
I = I_c\sin\phi + I_n(V).
\end{eqnarray}
Carriers in the nanowire remain normal, thus contributing to the instantaneous potential difference $V$. Since $V \sim d\phi/dt$, the observed time-averaged potential difference as a function of current is obtained by integrating the inverted equation
\begin{eqnarray}\label{eq:avV}
\frac{1}{\langle V \rangle} =
\frac{1}{2\pi} \int_0^{2\pi} \frac{d\phi}{V_n(\phi)} = \frac{1}{2\pi} \int_0^{2\pi} \frac{d\phi}{V_n(I-I_c\sin\phi)}.
\end{eqnarray}
Due to the presence of shunting normal component $I_n(V)$, or $V_n(I)$, Josephson oscillations do not average to zero in DC measurements, resulting in a wide current plateau terminated at $\pm I_c$ (see Figs.\,1b and 1c). From the geometry of the system, it is natural to model normal carrier transport by considering a system of two superconductor-normal interfaces separated by a metallic nanowire (due to the high doping levels). In this case, the normal current, $I_n$, can be found by calculating transmission and reflection processes at each interface\cite{BTK1982}. Due to significant scattering in the nanowire, carriers were allowed to re-thermalize (non-ballistic transport). The experimental data could be well described by fitting such a model (Fig.\,1c) except for the sub-gap region, where depairing contributions not included in the theoretical framework are observed. Furthermore, the fitting analysis indicated that one of the two interfaces was dominating, contributing most of the observed voltage drop. In an analogous manner, InAs nanowire Josephson junctions with significant disorder were previously described by a model relying on a lumped scatterer with a single effective transparency \cite{Abay2014}. Non-identical InAs/Al interfaces could also be explained by slight doping gradients along the nanowire axis or differences resulting from the nanofabrication process.

Four key parameters were extracted by fitting the IV curves as a function of temperature (Fig.\,1c) self-consistently: the critical current $I_c(T)$, the normal state resistance $R_n(T)$, the transmission coefficient $D$, and the effective superconducting gap of the Al leads at the interface with the nanowire.  Since our junctions are short and we do not expect the gap to vary through its length, this is also the "minigap" or proximity-induced gap inside the semiconductor, $2\Delta_m(T)$. To reduce the number of adjustable parameters per fit, we first extract the excess current $I_{ex}(T)$ and the normal state resistance $R_n(T)$ from the $IV$ curves above the sub-gap region $(V> 2\Delta$, in this case $V>125\,\mu V)$. The product of $I_{ex}$ and $R_n$ follows the expected BCS dependence as shown in Fig.\,1d. Having obtained $R_n(T)$ and assuming a temperature-independent transmission coefficient $D$ (resulting from the interface barrier height), the parameters $I_c(T)$, $\Delta_{m0} = \Delta_{m}(T=0)$, and $D$ are left to be determined. The critical current $I_c(T)$ is the only temperature dependent parameter left and was adjusted when fitting each of the IV curves obtained for different temperatures, while the transmission coefficient and $\Delta_{m0}$ were fixed for all temperatures, but adjusted over multiple fitting iterations. For the presented device, we obtained a transmission coefficient of 0.68 and $\Delta_0 = 60\,\mu eV$. Note that the gap energy is substantially smaller compared to the Al leads (208\,$\mu$eV), which were measured independently. This behaviour can be attributed to the inverse proximity effect and significant spin-orbit interaction in the nanowire material suppressing superconductivity, consistent with previous observations in similar devices\cite{Tiira2017}. The medium interface transparency of 0.68 can be attributed to the physical plasma etching employed in situ before Al deposition, potentially inducing disorder and scattering at the interfaces \cite{Gul2017}. Our nanofabrication procedure did not include a commonly used sulfur-based surface treatment\cite{Suyatin2007}; the physical etching approach was adopted due to its reliability, uniformity and reproducibility (five out of five nanowire junctions tested at cryogenic temperatures showed proximity-induced superconductivity.  See Supporting Information). The critical currents $I_c$ and the product critical current times normal-state resistance $I_cR_n$ extracted from the fits are shown in Fig.\,1e with a BCS-type temperature dependence for comparison. Neither $I_c$ nor $I_cR_n$ fit such a $\Delta(T)$ dependence (both decay more rapidly).  Furthermore, $I_c$ can not be fitted with ballistic transport model for Josephson junctions\cite{kuprianov1988}, (fit shown in Supporting Information), indicating the normal channel is diffusive.

\begin{figure}[H]
  \includegraphics[width=1\textwidth]{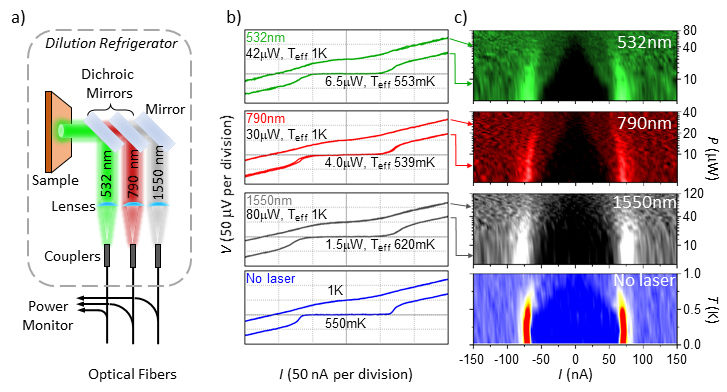}
  \caption{(a) Schematics of the experimental configuration used to characterize nanowire Josephson junctions under optical photon exposure.  (b) Three top graphs: IV curves at selected impinging laser powers corresponding to effective temperatures around $\sim$540-620\,mK and 1\,K for three different wavelengths. Bottom graph: IV curves obtained during sample stage heating experiments under no external illumination for comparison. (c) $dV/dI$ as a function of current bias showing the suppression of proximity-induced superconductivity for increasing impinging laser powers and sample stage temperatures.}
  \label{fgr2}
\end{figure}

The impact of photon exposure on the Josephson junction properties was studied under illumination for three different wavelengths using an optical setup inside the dilution refrigerator (Fig.\,2a). Optical fibers delivered the photons from three independent laser sources to the sample stage, where they were out-coupled to free space, collimated and directed onto the same position on the nanowire sample (only one laser was used at a time). The measured beam profile gave a Gaussian beam radius of 1.6\,mm, considerably larger than the devices under test. The transport properties of the nanowire Josephson junctions under constant current bias were characterized while the junctions were exposed to photons with wavelengths of 532\,nm, 790\,nm and 1550\,nm at increasing impinging laser power. Representative IV curves are shown in Fig.\,2b in comparison to the sample stage heating experiment.  The annotated effective temperatures were extracted using a fitting procedure detailed below. The qualitative behaviour for the sub-gap region of the IV curve at $V < 2\Delta$ was qualitatively similar for all four cases: proximity-induced superconductivity was gradually suppressed with increasing impinging laser power (temperature) and the IV curves began to adopt more linear Ohmic characteristics similar to the effect of heating. This is more clearly illustrated by color maps of the derivative $dV/dI$ (Fig.\,2c).  

\begin{figure}[ht]
  \includegraphics[width=1\textwidth]{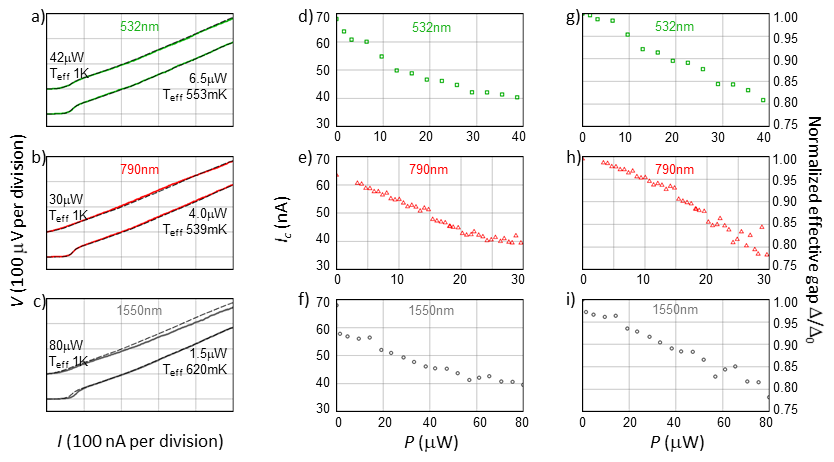}
  \caption{(a)-(c) IV curves at selected laser powers for three different wavelengths fitted with interpolated data obtained in sample stage heating experiments with no external illumination (dashed lines). The interpolated data was computed from IV curves recorded at different temperatures and was used to extract the effective temperature from the fit model. (d)-(f) The critical current $I_c$ deduced from the fit as a function of laser power. (g)-(i) The effective gap normalized to the value at base temperature under no laser illumination.}
  \label{fgr3}
\end{figure}

To quantitatively link the experimental results under photon exposure with the heating experiment, the sub-gap region of the IV curve ($V < 2\Delta$)  was fitted with temperature as only adjustable parameter. This was accomplished by first interpolating between the data points from the temperature experiment to arrive at an experimentally determined fitting function $V(I,T)$.  Then, each of the IV curves under laser illumination was fitted with this empirically obtained $V(I,T)$.  This empirical fit accounts for all thermally activated processes. Two exemplary fitting results for each wavelength are presented in Fig.\,3a-c. In the sub-gap region, the fit provided by the interpolation is in excellent agreement with experimental data, showing that laser illumination of the sample is equivalent to heating as far as the Josephson physics is concerned, and no new observable features are introduced. Therefore, we can assign an effective temperature $T_{eff}$ to each IV curve under illumination. $T_{eff}$ obtained from this procedure was then used in the model above to extract the critical current $I_c$ (Fig.\,3d-f) and the superconducting gap at the interface between the superconductor and the semiconductor (Fig.\,3g-i), as well as to establish a reliable relation between local sample temperature and the laser power for each wavelength. All three investigated wavelengths produced nearly identical heat dissipation pathways with $T_{\rm eff}\sim N^{1/3}$, where $N$ is the number of impinging photons (Fig.\,4a).

In contrast to the sub-gap region, the region of the IV curves above the gap, $V>2\Delta$, does change under illumination. The power dependence there is non-monotonic and does not map to any effective temperature dependence (see Fig.\,4b). Non-thermal effects in the normal state resistance under laser illumination can be attributed to the electron density affected by the number of absorbed photons, and to interface defects in the system being interacting with the incident light.  The complex fluctuation-like behavior is analogous to previous experimental observations on nanowire Josephson junctions under external gate voltage\cite{Guenel2012}. 

\begin{figure}[H]
  \includegraphics[width=1\textwidth]{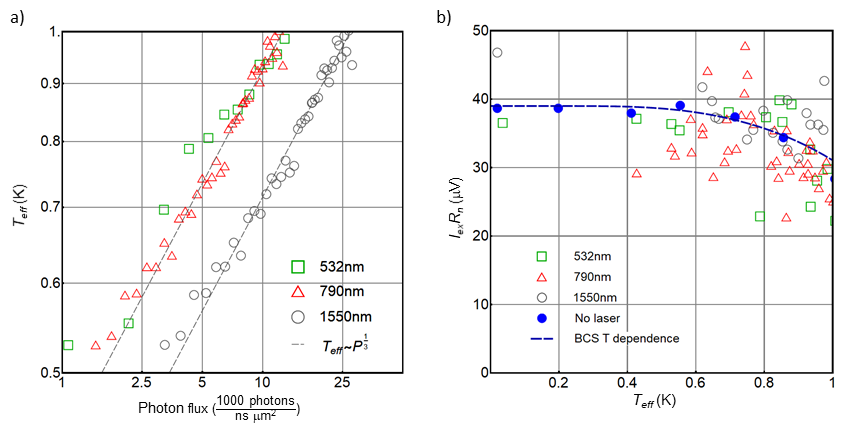}
  \caption{(a) Effective temperature $T_{eff}$ as a function of impinging photon flux for the three laser wavelengths. Dotted lines are guides to the eye, and reflect a $T\propto P^{1/3}$ relation. (b) The product of excess current times normal-state resistance $I_{ex}R_n$ as a function of effective temperature $T_{eff}$ for laser illumination at three different wavelengths and increasing sample stage temperatures (filled blue circles). BCS dependence (dashed line) describes the temperature data.  Non-monotonous deviations from BCS dependence were observed under photon exposure in contrast to the case of sample stage heating.}
  \label{fgr4}
\end{figure}

In conclusion, highly doped nanowire junctions are robust under illumination by optical photons; the effect of the incident photons on the Josephson physics can be effectively modeled using temperature as the sole parameter. On the other hand, the normal conducting part of the IV curves cannot be modeled using temperature alone.  It is surprising that while both the excess current (backed by Andreev reflection processes and superconductivity in the Al leads) and the combination $I_{ex}R_n$ exhibit similar non-monotonic changes with laser power as does normal state resistance, Josephson tunneling responsible for the superconducting plateau and $I_c$ appear to be unaffected by photon exposure, except via local temperature, suggesting a protection induced by the superconducting correlations. This invites further theoretical investigation of the self-shunted Josephson systems under illumination. The effective superconducting gap observed in these hybrid junctions begins to close as the effective temperature reaches approximately 0.6\,K, corresponding to incoming photon density of approximately 2$\times10^{19}$ - 4$\times10^{19}$ photons per second in the laser beam, or 2000-5000 photons per nanosecond per micrometer-square.  This offers a window in which a hybrid, highly-doped nanowire Josephson junction can be operated as an integral part of future hybrid superconducting optoelectronic circuits.

\begin{acknowledgement}
The fabrication and measurement at KTH are co-funded by Vinnova and Marie Curie Actions FP7-PEOPLE-2011-COFUND (GROWTH 291795), and the growth activity at NEST is co-funded by H2020-FETOPEN-2018-2020 (AndQC).
The authors would like to acknowledge Katharina Zeuner for her work related to the dilution refrigerator experimental setup and on maintaining the lasers used in this project.
\end{acknowledgement}

\begin{suppinfo}
Fabrication procedure; data from additional samples; data on the superconductivity of the Al leads under illumination; polarization dependence; ballistic junction fit to $I_c$.

\end{suppinfo}

\bibliography{references}

\newpage

\newpage
\section{Methods}
\subsubsection{Sample Fabrication}
The InAs/Al Josephson junctions were realized by means of a electron beam lithography lift-off process. The highly \textit{n}-doped InAs nanowires grown by chemical beam epitaxy were spun-cast on Si substrates with 150\,nm thermal oxide and pre-patterned Au alignment markers. Contact electrodes and bonding pads were aligned to the randomly positioned nanowires using scanning electron microscopy images of the nanowires and their adjacent alignment markers. Electron beam lithography was performed on a bi-layer PMMA resist (approximately 200\,nm for each layer) by a Raith Voyager system (50\,kV electron acceleration voltage). Automatic writefield alignment procedures were employed to ensure precise positioning of the Al electrodes on the InAs nanowires. After resist development the samples were first cleaned with a mild oxygen plasma for 15\,s to remove residual resist.  The plasma was kept short to avoid additionally oxidizing the nanowire surface.  Next, the sample was loaded into an AJA Orion magnetron sputtering tool. The native oxide layer on the nanowire surfaces was removed in situ with Ar ions in a physical plasma etching process (15\,mTorr, 45\,W RF power, 3\,min excluding power ramping). Subsequently, an Al layer of 100\,nm was deposited at 150\,W RF power and 3\,mTorr Ar pressure.  To facilitate lift-off, the sample holder was not rotated during to deposition.  After deposition, the samples were immersed in standard resist remover overnight at room temperature.  The majority of the Al is then mechanically removed from the surface by tweezers, and ultrasonication in a clean beaker of remover at room temperature was employed to complete lift-off.

\subsubsection{Transport Measurements}
 The electrical transport properties were characterized in a Bluefors dilution refrigerator with a base temperature around 10\,mK. The InAs/Al Josephson junction devices were wire-bonded to ceramic chip packages and mounted on the mixing chamber stage, being electrically connected to thermally anchored low-pass RC filters. Two RC filters were connected in series for each electrical line.  Outside the dilution refrigerator, the electrical lines are filtered with pi-filters at room temperature.  IV measurements were performed in a four-point configuration with a customized battery-powered instrument (IVVI rack developed at TU Delft) for current biasing and voltage readout. The experiments under light illumination employed commercially available optical components (ThorLabs) and continuous-wave laser sources at the three investigated wavelengths.
 
\newpage
\section{Supporting Figures}
Five Josephson devices were fabricated and three were investigated under laser illumination.  Device A refers to the device presented in the main text.  All junctions showed similar behavior.  Figures 1 and 2 below summarize results from two other nanowires.  
\begin{figure}[h!]
  \centering
  \includegraphics[width=1
  \textwidth]{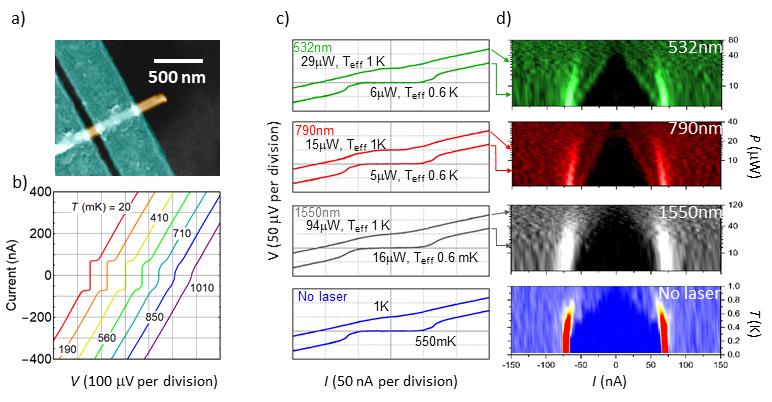}
   \caption{Device B: Nanowire Josephson junction at a different location on the same chip.  a) SEM image of device.  b) IV characteristics obtained under DC current biasing at different cryostat temperatures (the sample was not exposed to photons during the measurements). c) Representative IV curves at different illumination powers.  For IV under illumination power, the curves were chosen for effective temperatures 600\,mK and 1\,K.  d) Color map of dV/dI(I) for different illumination powers and at different temperatures, showing suppression of supercurrent.}
  \label{figrS1}
\end{figure}

\begin{figure}[H]
  \includegraphics[width=.8\textwidth]{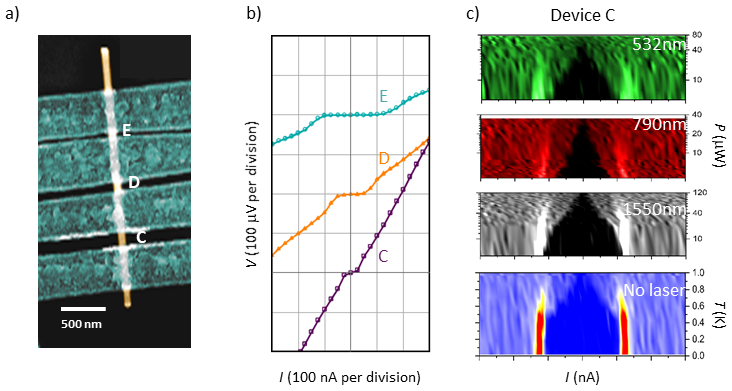}
  \caption{a) SEM image of nanowire with three Josephson junctions.  b) $IV$curves at base temperature without laser illumination.  The widest junction shows the largest normal state resistance and the lowest $I_c$.  c) The behavior of the $IV$ curves as functions of temperature and light exposure for junction C.}
  \label{fgrS2}
\end{figure}

For each Josephson device, we also measured the $IV$ through one of the Al electrodes as functions of substrate temperature and laser illumination power.  The Al was found to be a standard superconductor obeying a BCS dependence both as a function of temperature and incident laser power.  An exemplary series of IV curves of the Al leads is shown here for the device reported in the main paper (Fig. 3).  

\begin{figure}[H]
  \includegraphics[width=.7\textwidth]{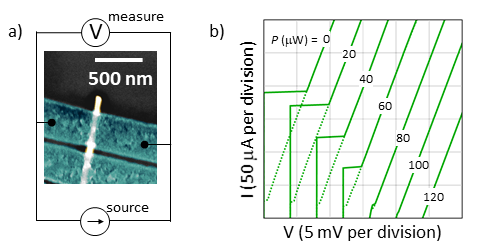}
  \caption{a) SEM image of device A (device presented in main text).  The four-point measurement configuration used to measure the outermost Al contact is schematically indicated. b) Seven exemplary IV curves of the Al lead under 532\,nm laser illumination. Exposure to photons of 790\,nm and 1550\,nm wavelength produced qualitatively the same effect.}
  \label{fgrS3}
\end{figure}

Due to their geometry, the optical absorption of bare semiconductor nanowires is polarization dependent.  However, the exposed region of the nanowire in our Josephson junction devices is symmetric (nanowire diameter is approximately equal to the gap $L$ between the Al leads).  We used the laser at 790\,nm to verify that the response of our Josephson junctions is insensitive to the polarization of the incident light as follows:  We replaced the power-meter on the monitoring arm of the beamsplitter outside of the dilution refrigerator by a polarometer.  The laser was set at an intermediate power, 10\,$\mu W$, at which its effect on the $IV$ curve was clearly noticeable during the power series. The polarization on the input arm  of the beamsplitter was set with a 2-paddle polarization controller.  Lacking polarization optics and polarization maintaining fibers inside the dilution refrigerator, we cannot determine the polarization of the light relative to our devices, however, by setting the input poarization to all six orthogonal polarization states (horizontal, vertical, diagonal, anti-diagonal, left-circular, and right-circular), we can determine whether there is any effect due to polarization of the incident light. The resulting $IV$ curves are clearly insensitive to polarization, as shown in Fig. S3 below.  In contrast, the $IV$ curves of the Al leads are sensitive to incident light polarization, with $I_c$ varying between 15$\mu A$ to 75\,$\mu A$. 

\begin{figure}[H]
  \includegraphics[width=0.7
  \textwidth]{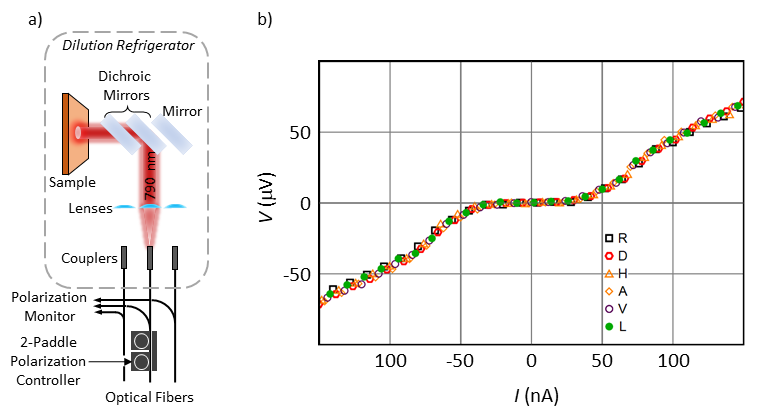}
  \caption{Polarization dependence of the device shown in main text.  a) Schematic of the setup: Laser wavelength 790\,nm,  10\,$\mu$W, polarization was monitored outside of the cryostat. b) Six IV curves, one for each incident polarization, are almost identical, verifying that the device characteristics are insensitive to the polarization of the incident light.}
  \label{fgrS4}
\end{figure}

\newpage
The evolution of $I_c$ as a function of stage temperature (no laser illumination) cannot be fitted by the analytical expression derived by Kuprianov and Lukichev (Reference 35 in main text). Furthermore, the temperature dependence trend of $I_c$ is similar to those derived for diffusive junctions (requiring a numerical fit).  We conclude that transport in our junctions is diffusive.
\begin{figure}[H]
  \includegraphics[width=0.5
  \textwidth]{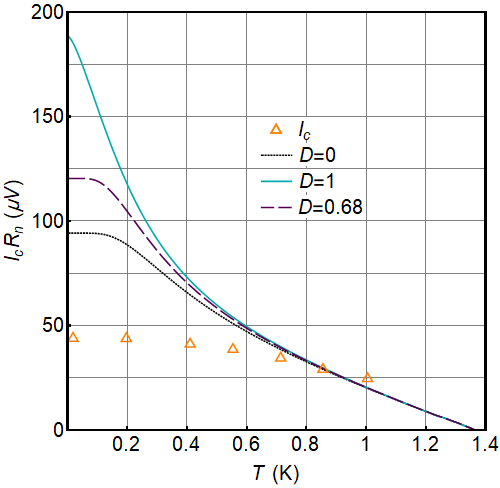}
  \caption{Experimental data (orange triangle) and three theoretical curves based on analytical formula derived in Reference 35 in main text.  Black and Blue lines are the two limiting cases with $D=0$ and $D=1$, repectively.  Dotted purple line reflects our device with $D=0.68$.  No fitting parameters could be found to fit $I_c$ vs T for our data.}
  \label{fgrS5}
\end{figure}


\end{document}